\documentclass[a4paper,11pt]{article}
\usepackage{amsmath}
\usepackage{amssymb}
\usepackage{graphicx}
\usepackage[sort]{cite}
\usepackage[T1]{fontenc}
\usepackage[a4paper,left=1.0in, right=1.0in,top=1.1in, bottom=1.2in]{geometry}
\usepackage[affil-it]{authblk}
\usepackage{color}
\newcommand{\be}{\begin{equation}}
\newcommand{\ee}{\end{equation}}
\newcommand{\pythia}{\textsc{Pythia}}

\newcommand{\pt}{p_{\rm T}}  
\newcommand{\xf}{x_{\rm F}}
\renewcommand{\d}{{\rm d}}
\newcommand{\pthat}{\hat{p}_{\rm T}}

\begin{document}

\title{\Large \bf Single inclusive jet transverse momentum and energy spectra\\ at very forward rapidity in proton-proton collisions\\ with $\sqrt{s}$ = 7 and 13 TeV}

\author{Krzysztof Kutak}
\affil{Instytut Fizyki Jadrowej, Radzikowskiego 152, 31-342 Krak\'ow, Poland}
\author{Hans Van Haevermaet and Pierre Van Mechelen}
\affil{Particle Physics group, University of Antwerp, Belgium}

\maketitle

\begin{abstract}
We present predictions of single inclusive jet transverse momentum, energy, and Feynman-$x$ spectra at forward rapidity ($5.2 < y < 6.6$) in proton-proton collisions with $\sqrt{s} = 7$ and 13 TeV.  Calculations based on high-energy factorization and $k_T$-dependent parton densities are compared to simulations using the \pythia\ event generator.  Effects from parton density evolution, parton shower dynamics, multi-parton interactions, and saturation of parton densities are investigated.  
\end{abstract}

\section{Introduction}

The study of single inclusive forward jet production allows to investigate various aspects of hadron-hadron scattering.  Jets resulting from a hard parton interaction are boosted forward if the incoming partons have a large imbalance in the fractional hadron's momenta, $x$, carried by the partons. Such processes are therefore ideal to test approaches that allow for studies of both high-$x$ and low-$x$ phenomena. Moreover, at large rapidity, the transverse momentum of the jet is kinematically bound to small values, making this process very sensitive to the modeling of the underlying event (i.e.\@ initial and final state parton showers, multi-parton interactions, and beam remnant fragmentation), regularization of the partonic cross section, and (perturbative) saturation \cite{Gribov:1984tu} of parton densities. 

In particular this last phenomenon, i.e.\@ the saturation of gluons in hadrons,  is one of the open problems in QCD. It is related to perturbative unitarity of the QCD evolution equations and follows from constraints on the rate of growth of the cross section as the energy of the collision increases.  Microscopically, saturation is an outcome of the competition between gluon splitting and gluon fusion processes, and can be theoretically described by nonlinear QCD evolution equations \cite{Balitsky:1995ub,Kovchegov:1999yj,Kovchegov:1999ua,JalilianMarian:1997gr,Iancu:2001ad}. Phenomenological studies of various processes are compatible with the existence of saturation in Nature \cite{GolecBiernat:1998js,Albacete:2010pg,Dumitru:2010iy,Dusling:2013qoz,Kutak:2012rf}.

In this paper we discuss some of the open questions raised above, and we present predictions of single inclusive jet transverse momentum, energy, and Feynman-$x$ spectra with rapidity $5.2 < y < 6.6$ in proton-proton collisions at $\sqrt{s} = 7$ and 13 TeV.  The chosen rapidity range corresponds to the acceptance of the CASTOR calorimeter installed at the CMS experiment \cite{Chatrchyan:2008aa}, which has collected data at the LHC with pp collisions at various center-of-mass energies.  

The spectra are calculated using high-energy factorization (HEF) \cite{Catani:1990eg,Deak:2009xt} and $k_T$-dependent parton densities. In this approach, matrix elements for single inclusive jet production can be given at leading order as a $2 \to 1$ process with one of the incoming partons being off-shell. This is in contrast to collinear factorization, where the matrix element for the $2 \to 1$ process vanishes at leading order, and where one has to include corrections at higher order in $\alpha_{\rm S}$ to account for the finite transverse momentum of the jet.  

The results of this calculation are compared to predictions obtained with the \pythia\ Monte Carlo program \cite{Sjostrand:2014zea,Sjostrand:2006za}, which is based on collinear factorization.  In addition, the \pythia\ event generator also includes models for the underlying event, and the regularization of the partonic cross section.

\section{Single inclusive jet production in high-energy factorization}

The single inclusive jet production process can be schematically written as
\begin{equation}
{\rm A} + {\rm B} \rightarrow {\rm a} + {\rm b} \rightarrow \text{jet} + {\rm X}
\end{equation}
where A and B are the colliding hadrons, each of which provides a parton, respectively a
and b, and X corresponds to undetected, real radiation. The beam remnants from the hadrons A
and B are understood to be implicitly included in the above equation.

The longitudinal kinematic variables can be expressed as
\begin{equation}
x_1 = \frac{1}{\sqrt{s}}\, p_{\rm T}\, e^{y}, \qquad x_2 = \frac{1}{\sqrt{s}}\, p_{\rm T}\, e^{-y},
\end{equation}
with $s=(p_{\rm A}+p_{\rm B})^2$  the total squared energy of the colliding hadrons, while $y$ and $p_{\rm T}$ are the rapidity and transverse momentum of the leading final state jet, respectively.

The HEF\footnote{The formula is at leading order accuracy. There is ongoing activity to advance it to NLO level \cite{Iancu:2016vyg,Chirilli:2011km,Altinoluk:2014eka}. For a review see \cite{Stasto:2016wrf}, and for an overview of applications of HEF framework to other processes see 
\cite{Deak:2010gk,Deak:2011ga,Deak:2009ae,Deak:2011gj,Kutak:2016mik,Kutak:2016ukc,Luszczak:2016csq,Baranov:2015yea,Dooling:2014kia,Sapeta:2015gee} and references therein.} formula applicable for the rapidity range that we address in this paper reads \cite{Dumitru:2005gt}:
\begin{equation}
\begin{split}
\frac{{\rm d}\sigma}{{\rm d}y\, {\rm d}p_{\rm T}} = \frac{1}{2}\frac{\pi\, p_{\rm T}}{(x_1 x_2 s)^2} & \bigg[\sum_{\rm q(\bar q)}\overline{|{\cal M}_{\rm g^*q(\bar q)\to q (\bar q)}|}^2 x_1 f_{\rm q(\bar q)/A}(x_1,\mu^2)\, {\cal F}_{\rm g^*/B}^{\rm F}(x_2,p_{\rm T}^2,\mu^2)\\
&+\overline{|{\cal M}_{\rm g^*g\to g}|}^2 x_1 g_{\rm g/A}(x_1,\mu^2)\, {\cal F}_{\rm g^*/B}^{\rm A}(x_2,p_{\rm T}^2,\mu^2))\bigg] \,.
   \label{eq:int-phi}
\end{split}
\end{equation}
where ${\cal F}^{\rm F}$ is the unintegrated gluon density in color fundamental representation, while  ${\cal F}^{\rm A}$
is the unintegrated gluon density in color adjoint representation. These functions depend on the longitudinal momentum fraction $x$ and transverse momentum $\pt$, and in general also on the hard scale $\mu$. The matrix elements squared represent the scattering of an off-shell gluon with an on-shell quark, and an off-shell gluon with an on-shell gluon, respectively, and are averaged over initial state helicity (indicated by the bar) and summed over final state helicity. They can be obtained directly by application of helicity methods \cite{vanHameren:2012if, Bury:2016cue}. For the observables that we study here, it is known that the hard scale dependence of the unintegrated gluon density does not affect the cross section \cite{Bury:2016cue}.  

We use unintegrated gluon densities which follow from the extended BFKL \cite{Kuraev:1977fs,Balitsky:1978ic,Kuraev:1976ge} (yielding the KS linear gluon density) and BK (yielding the KS nonlinear gluon density) evolution equations \cite{Balitsky:1995ub,Kovchegov:1999yj,Kutak:2012rf}. The latter gluon density includes effects from saturation due to the contribution of nonlinear terms. 

The formula above can be used to construct the following observables:
\begin{itemize}
\item Transverse momentum spectrum:
\begin{equation}
\frac{{\rm d}\sigma}{{\rm d}p_t}= \int_{y_{\rm min}}^{y^{\rm max}} \d y\, \frac{{\rm d}\sigma}{{\rm d}y\, {\rm d}p_{\rm T}} 
\end{equation}
\item Energy spectrum:
\begin{equation}
\frac{\d\sigma}{\d E}= \int_{y_{\rm min}}^{y_{\rm max}} \d y\, \frac{\d\sigma}{\d y\,\d p_{\rm T}}\frac{1}{\cosh\,y} 
\label{eq:energy}
\end{equation}
\item Feynman-$x$ spectrum ($x_{\rm F} =  2\, p_{\rm L}/\sqrt{s}$, with $p_{\rm L}$ the longitudinal momentum): 
\begin{equation}
\frac{\d\sigma}{\d x_{\rm F}}= \frac{\sqrt{s}}{2} \frac{\d \sigma}{\d p_{\rm L}} = \frac{\sqrt{s}}{2} \int_{y_{\rm min}}^{y_{\rm max}} \d y\, \frac{\d\sigma}{\d y\, \d p_{\rm T}} \frac{1}{\sinh y} 
\end{equation}
\end{itemize}
Partons are assumed to be massless, so that rapidity equals pseudorapidity.  Here, we take $y_{\rm min} = 5.2$ and $y_{\rm max} = 6.6$, which corresponds to the acceptance of the CASTOR calorimeter.  Note that for the rapidity values that are considered, $\cosh y \approx \sinh y$, and $x_{\rm F}$ can be approximated by $2\, E/\sqrt{s}$.

\section{Single inclusive jet production with \pythia}

The \pythia\ Monte Carlo event generator (version 8.212) is used to study forward jet production in the collinear factorization framework.   

In a first, basic configuration, $2 \to 2$ QCD processes (${\rm gg \to gg}$ and ${\rm qg \to qg}$) above a certain threshold of exchanged transverse momentum, $\pthat$, are generated and matched with initial state radiation (ISR) parton showers (labeled ``Hard QCD + ISR''), which are transverse-momentum-ordered and governed by the DGLAP evolution equations \cite{Sjostrand:2014zea}. Other physical effects, such as final state radiation (FSR) and hadronization are switched off for the generation of this event sample. The partonic $2\to 2$ cross section for these processes diverges as $1/\pthat^{4}$, and would lead to an unphysical total inelastic cross section if no further regularization is applied.  To illustrate this behavior, samples with a minimal transverse momentum of $\hat{p}_{\rm T,min}$ = 0.5 GeV and $\hat{p}_{\rm T,min}$ = 2.0 GeV are obtained, both with the minimum allowed invariant mass of the outgoing two-parton system set to zero.

A more realistic configuration of \pythia\ is provided by the Monash tune \cite{Skands:2014pea}. Here, so-called soft QCD events are generated that contain a mix of semi-hard $2 \to2$ QCD processes in an eikonalized description, intended to be valid at all $\pthat$.  An important element of this setup is the regularization of the cross section by applying a smooth dampening factor, $1/\pthat^{4} \to 1/(\pthat^2 + \hat{p}_{\rm T,0}^2)^2$, to the partonic cross section and the introduction of multiple parton interactions (MPI).  In order to study the influence of FSR and hadronization separately, samples are generated with only ISR (labeled ``Soft QCD + ISR''), with ISR and FSR (``Soft QCD + ISR, FSR''), and with ISR, FSR, and hadronization (``Soft QCD + ISR, FSR, hadr.'').

In all MC samples, jets are clustered with FastJet \cite{Cacciari:2011ma} using the anti-$k_{\rm T}$ algorithm with distance parameter equal to 0.5 \cite{Cacciari:2008gp}.

\section{Numerical results}

\subsection{Energy and transverse momentum spectra}

\begin{figure}[t]
\centering
\includegraphics[width=0.49\textwidth]{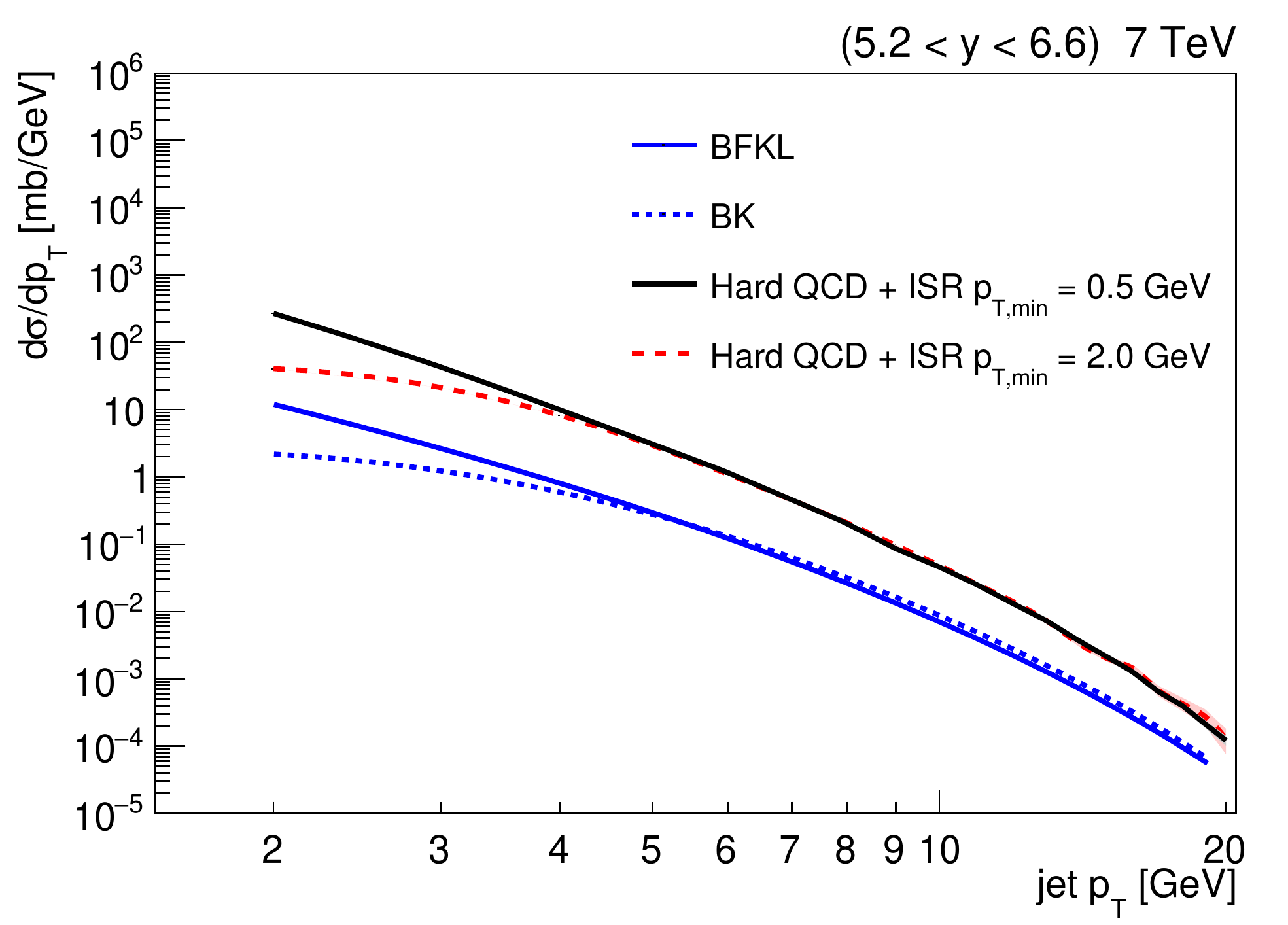}
\includegraphics[width=0.49\textwidth]{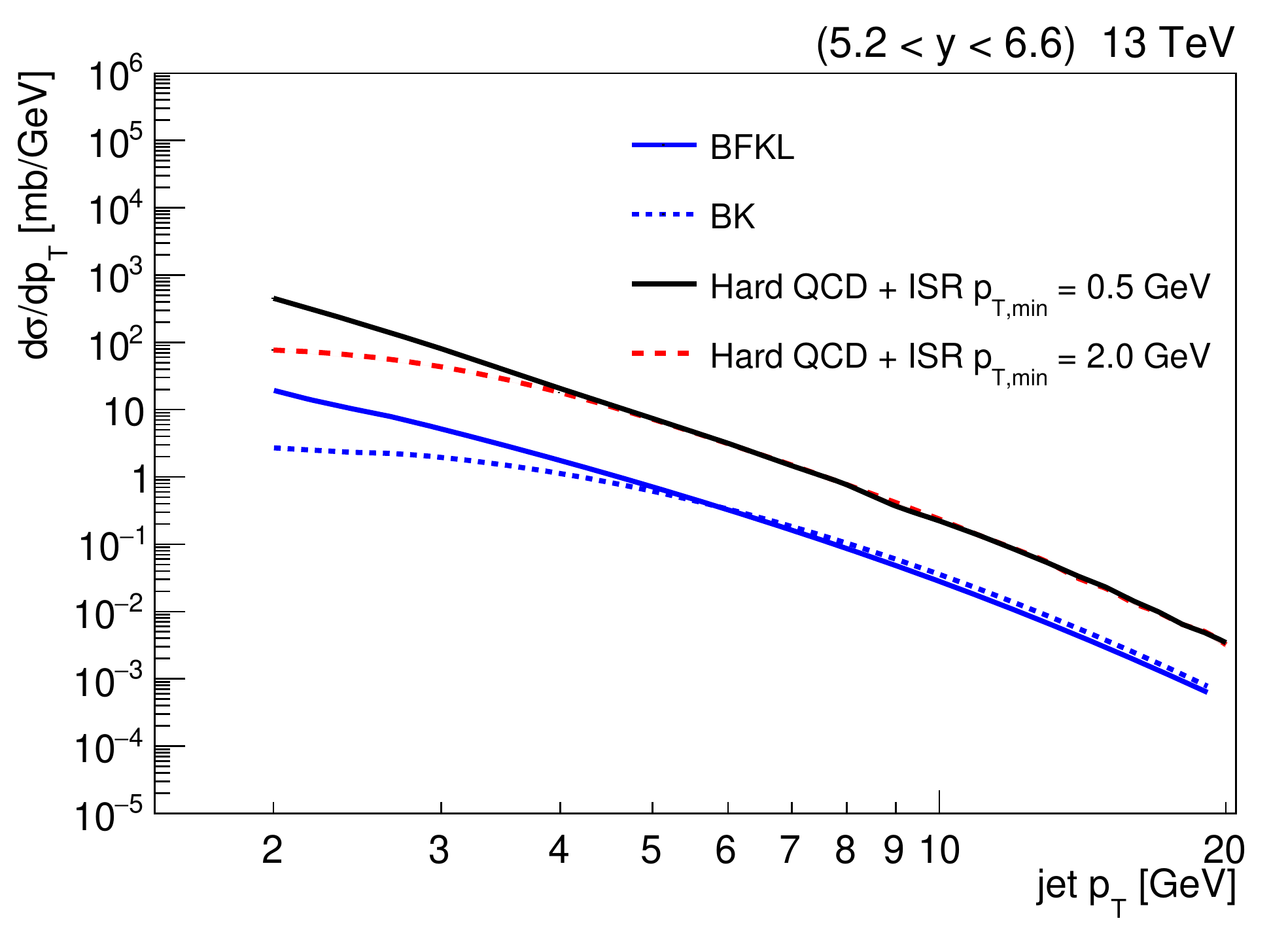}
\caption{Differential jet cross sections as function of jet $\pt$ at forward rapidity $5.2 < y < 6.6$ and for $\sqrt{s} = 7$ TeV (left) and 13 TeV (right).  Calculations obtained with the HEF framework (labeled ``BFKL'' and ``BK'') are compared to simulations obtained with \pythia\ for hard QCD processes with ISR and two different lower cutoffs on $\pthat$, the transverse momentum scale of the hard subprocess.}
\label{fig:pTspectra_hardQCD}
\end{figure}

The single inclusive jet transverse momentum spectrum is shown in Figure \ref{fig:pTspectra_hardQCD}
for center-of-mass energies $\sqrt{s} = 7$ and 13 TeV. Calculations with the HEF framework are compared to simulations using the \pythia\ Monte Carlo event generator for ``Hard QCD'' processes with only ISR included.

The HEF calculations predict a dramatic suppression of the low energy part of the spectrum with the nonlinear parton densities w.r.t.\@ the linear ones. We attribute this effect to saturation of the unintegrated gluon density, which is indeed expected to manifest itself at low $x$ and $\pt$.  In the HEF framework with one off-shell initial state parton, the $\pt$ of the produced jet directly corresponds to the transverse momentum of the incoming parton.  When one lowers the jet $\pt$, the sensitivity to the saturation scale becomes more visible. Moreover, in a fixed rapidity window, the fractional momentum $x$ that is probed is also smaller for small jet $\pt$.  The region of high jet $\pt$ probes the parton densities at a scale that is much larger than the saturation scale and the predictions based on the linear (BFKL) and nonlinear (BK) evolution equations are therefore consistent, with both exhibiting a power law dependence.

The HEF calculation does not include final state radiation and hadronization effects, while initial state radiation is taken into account via the $\pt$ dependence of parton densities, as an outcome of the evolution in rapidity of the unintegrated parton density.  It has been argued that MPI effects are also at least partially included in HEF \cite{Jung,Kotko:2016lej}.   We therefore compare with the \pythia\ predictions for hard QCD $2\to 2$ processes with final state radiation and hadronization effects turned off. MPI processes are also not included in this configuration.  The \pythia\ predictions are remarkably parallel to the HEF calculations, and are reminiscent of the $1/\pthat^{4}$ dependence of the partonic cross section. There exists however an important difference in normalization of the two calculations.  It should be noted that the total inelastic cross section reported by \pythia\ is compatible with measurements only after all effects discussed below (soft regularization, MPI, etc.) are included.  We observe that introducing a lower cutoff of 2 GeV on $\pthat$ has a very similar effect as using the nonlinear KS gluon density.

The qualitative agreement in the small and moderate transverse momentum domain of the HEF prediction with \pythia\ is consistent with the observation already made in \cite{Bury:2016cue}, and points at the consistency between the considered frameworks in re-summation of logarithms of transverse momenta when the momenta are moderate.  

In general one could expect that the BFKL or BK approach should provide relatively more high energy jets than the collinear framework.  However, the results show that the predictions are rather comparable in shape. This is due to the inclusion of higher orders in the BFKL and BK evolution following the KMS prescription \cite{Kwiecinski:1997ee}:  kinematic constraints and complete splitting functions that limit the phase space for energetic emissions render the result similar to collinear factorization with initial state parton showers as implemented in \pythia. 

\begin{figure}[t]
\centering
\includegraphics[width=0.49\textwidth]{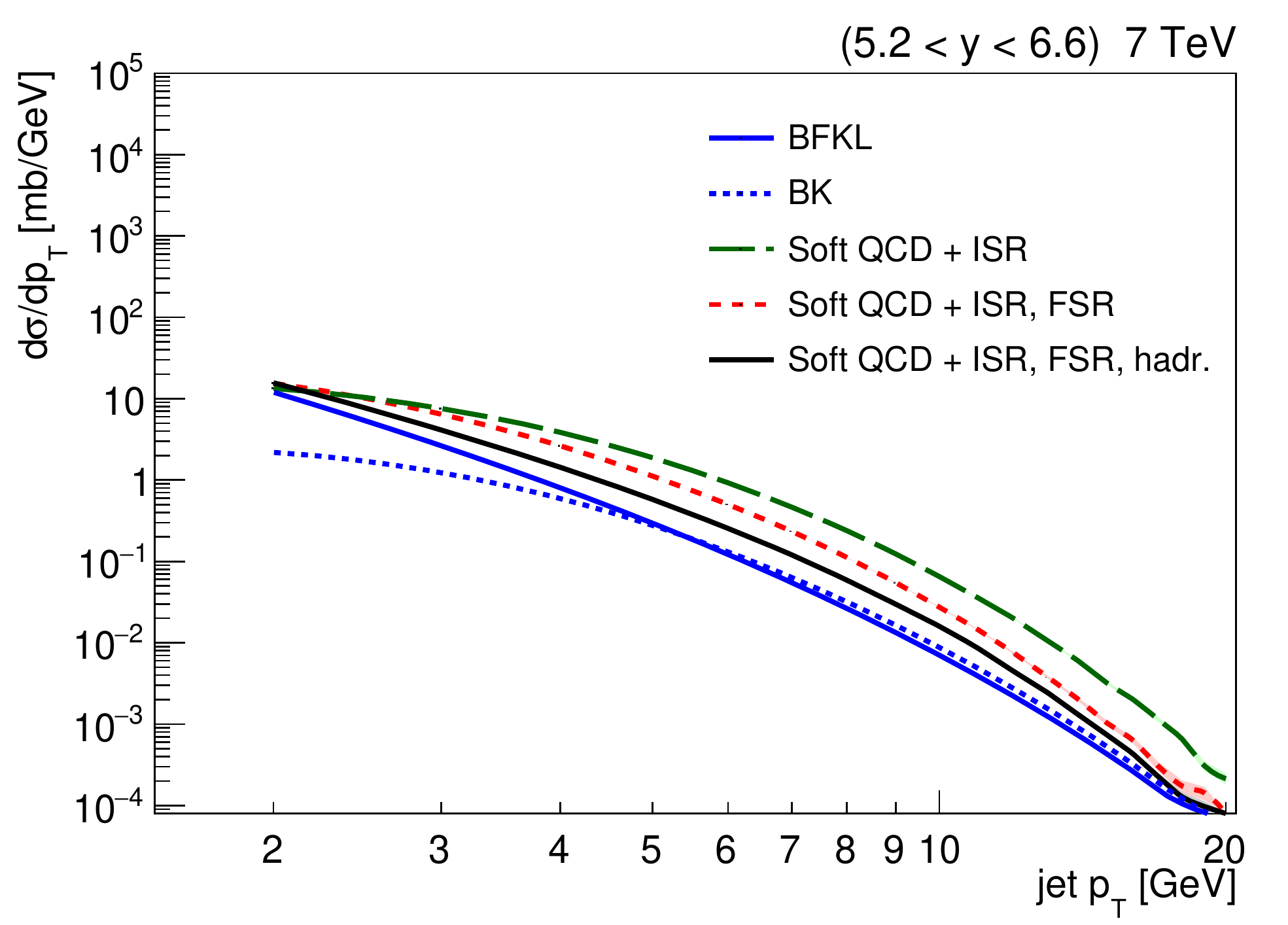}
\includegraphics[width=0.49\textwidth]{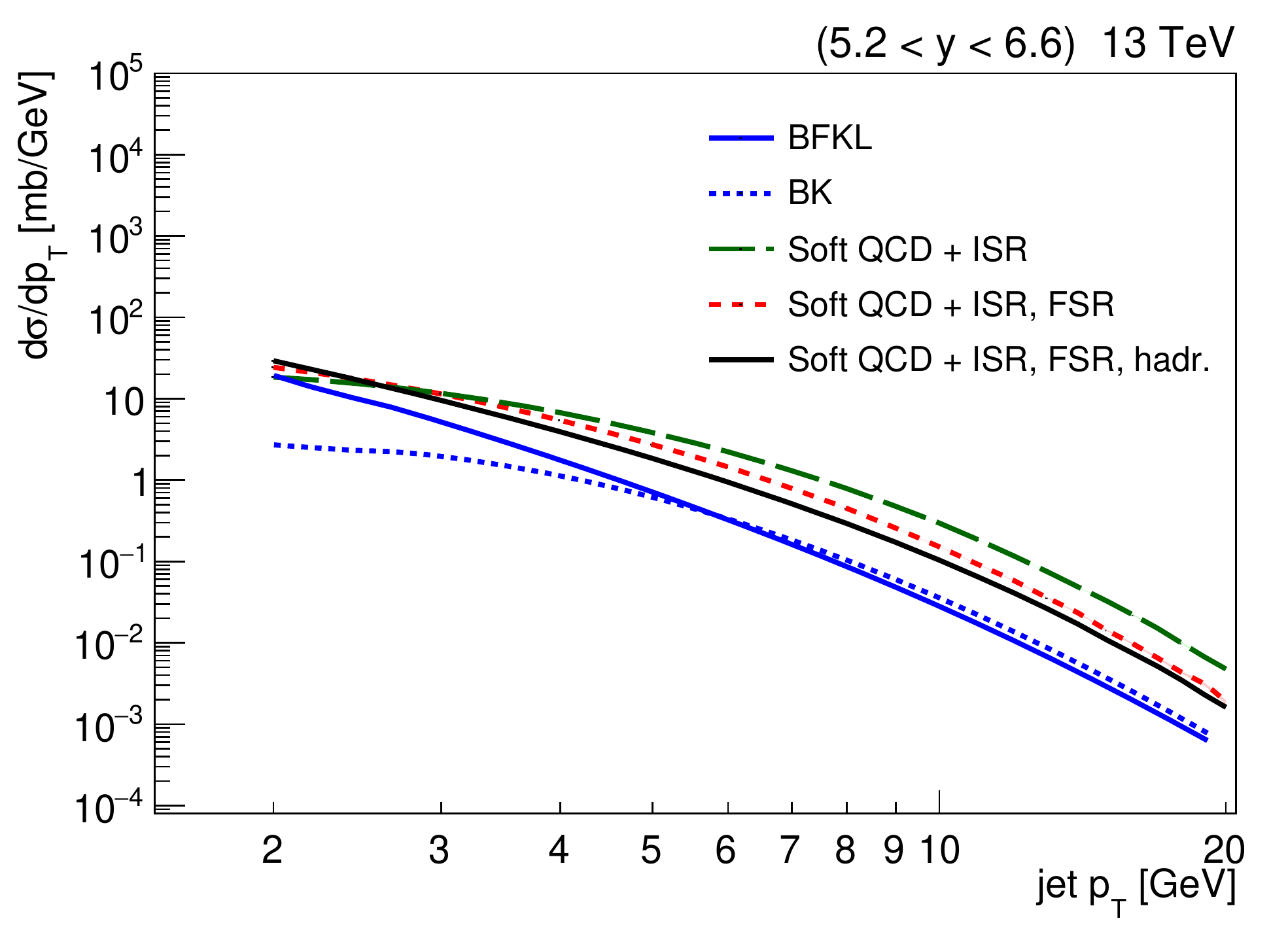}                
\caption{Differential jet cross sections as function of jet $\pt$ at forward rapidity $5.2 < y < 6.6$ and for $\sqrt{s} = 7$ TeV (left) and 13 TeV (right).  Calculations obtained with the HEF framework (labeled ``BFKL'' and ``BK'') are compared to simulations obtained with \pythia\ for soft QCD processes with ISR, adding subsequently FSR and hadronization.}
\label{fig:pTspectra_softQCD}
\end{figure}

A more realistic description of low-$p_{\rm T}$ QCD processes is available in \pythia\ via the generation of ``soft QCD'' collisions, including a smooth regularization of the partonic cross section and MPIs.  Figure~\ref{fig:pTspectra_softQCD} shows a comparison of the $\pt$ spectra predicted by the HEF framework and soft QCD events in \pythia.

The shape of the $\pt$ spectrum for soft QCD events is drastically different, and the resulting total inelastic cross section is much more compatible with measurements.  Figure \ref{fig:pTspectra_softQCD} also shows the effect of final state radiation and hadronization as modeled in \pythia. Adding FSR to the simulation results in a suppression of the high energy tail of the spectrum because energy is radiated outside the jet cone. Adding hadronization effects softens the spectrum even further, which can be explained by the presence of softer particles originating from the fragmentation of partons.  The effect of hadronization is slightly stronger at $\sqrt{s} = 7$ TeV than at $\sqrt{s} = 13$ TeV, especially for the energy and $\xf$ spectra discussed below.

\begin{figure}[t]
\centering
\includegraphics[width=0.49\textwidth]{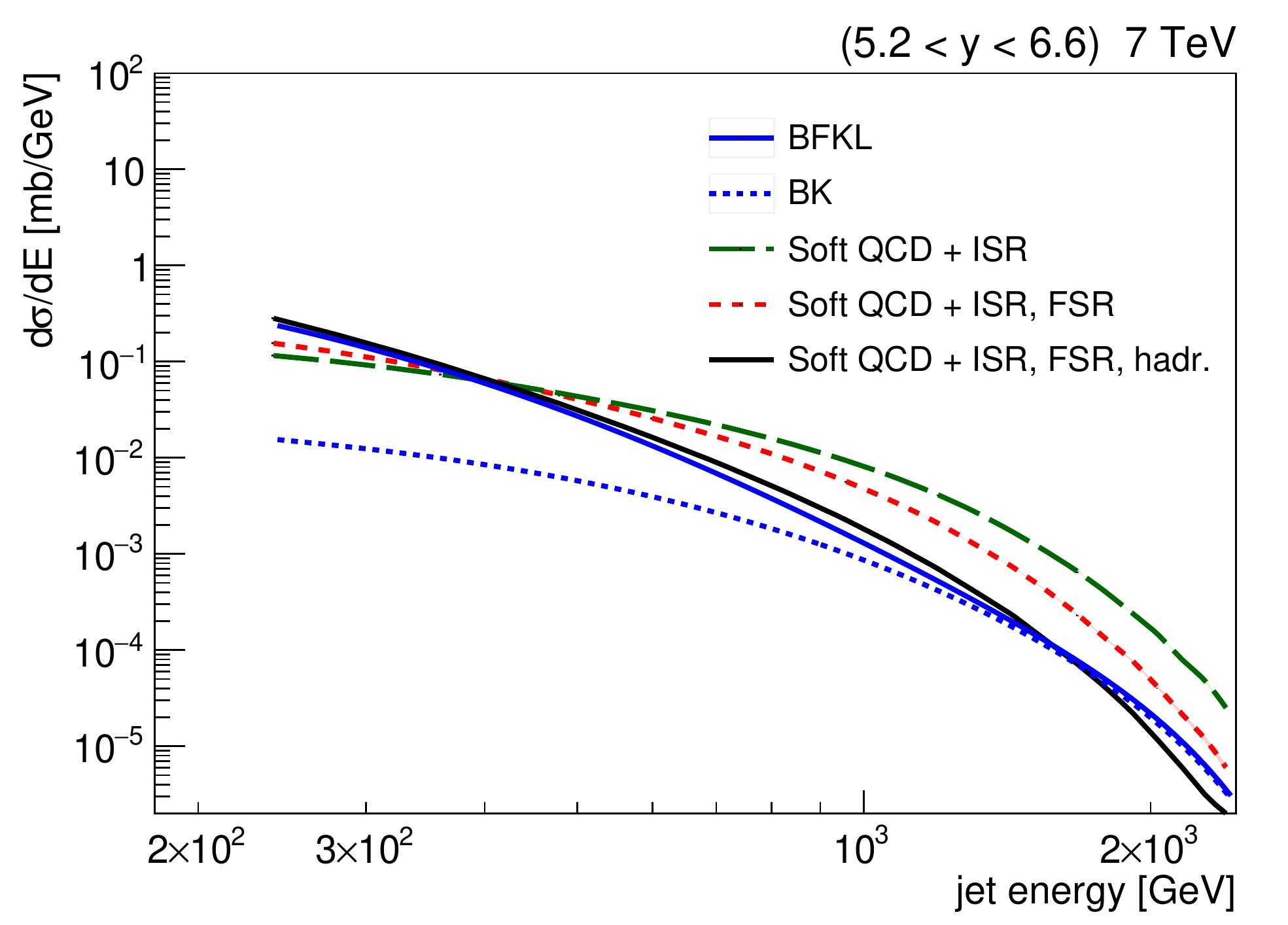}
\includegraphics[width=0.49\textwidth]{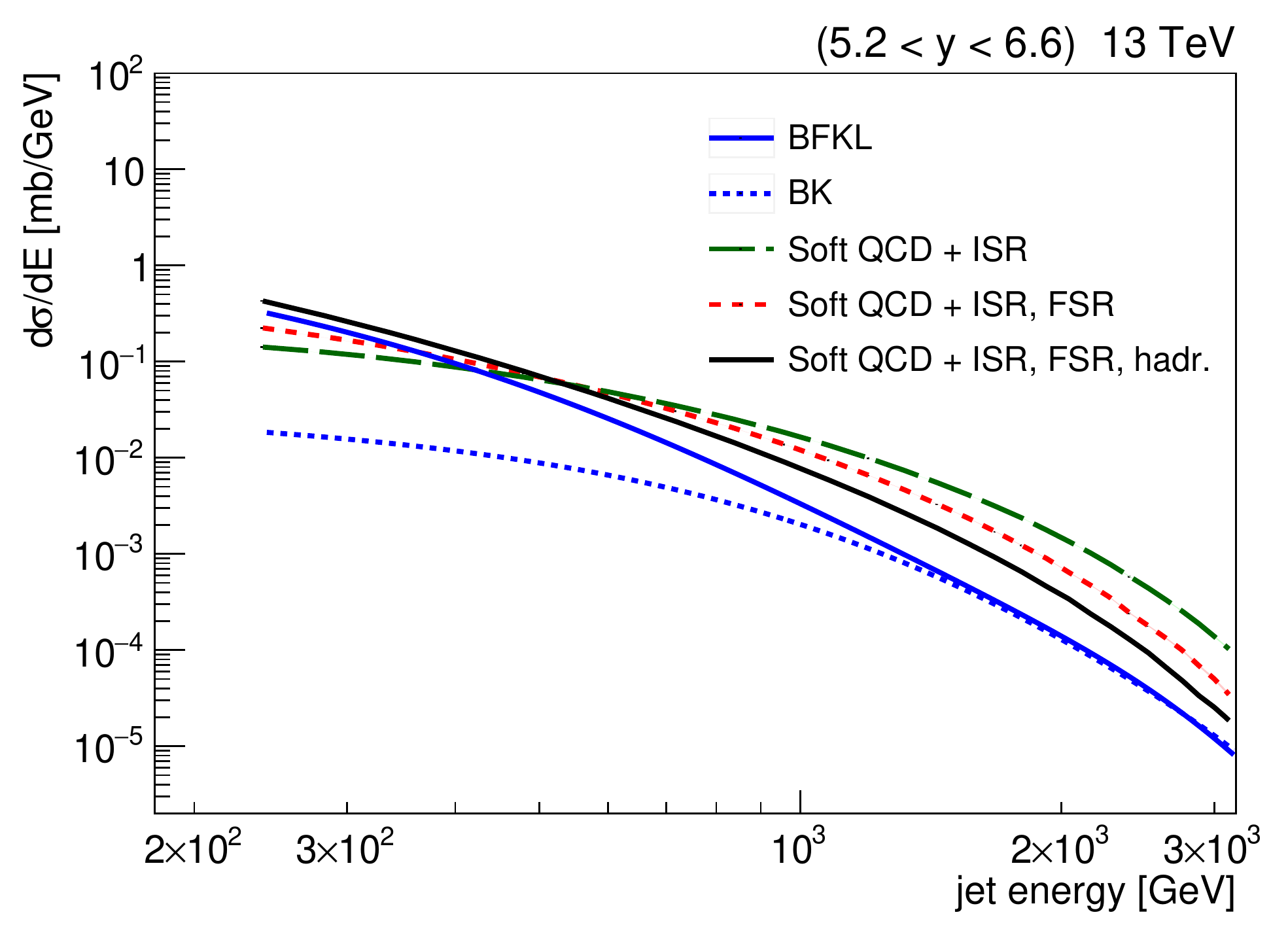}                
\caption{Differential jet cross sections as function of jet energy at forward rapidity $5.2 < y < 6.6$ and for $\sqrt{s} = 7$ TeV (left) and 13 TeV (right).  Calculations obtained with the HEF framework (labeled ``BFKL'' and ``BK'') are compared to simulations obtained with \pythia\ for soft QCD processes with ISR, adding subsequently FSR and hadronization.}
\label{fig:Energyspectra}
\end{figure}

\begin{figure}[t]
\centering
\includegraphics[width=0.49\textwidth]{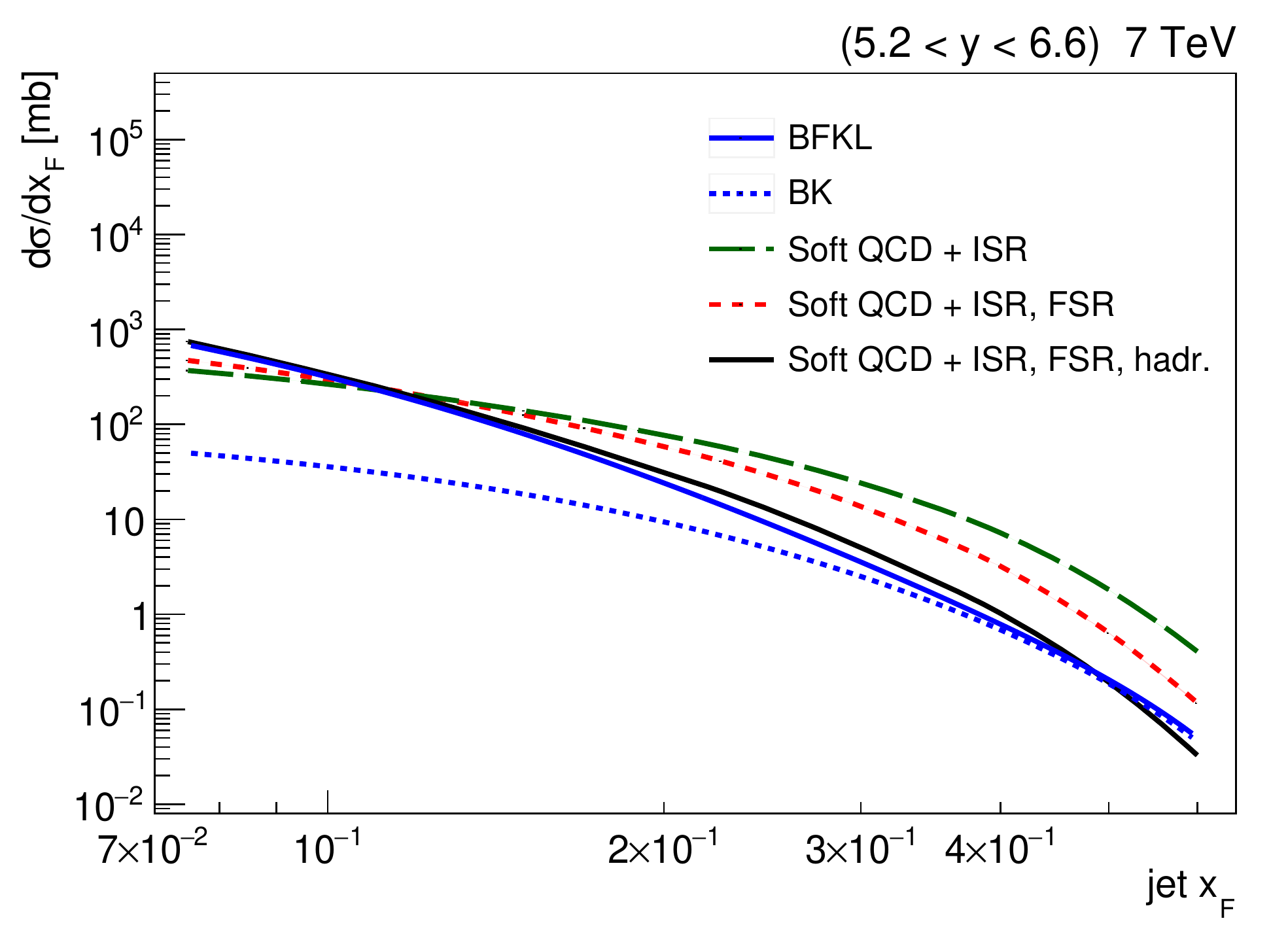}
\includegraphics[width=0.49\textwidth]{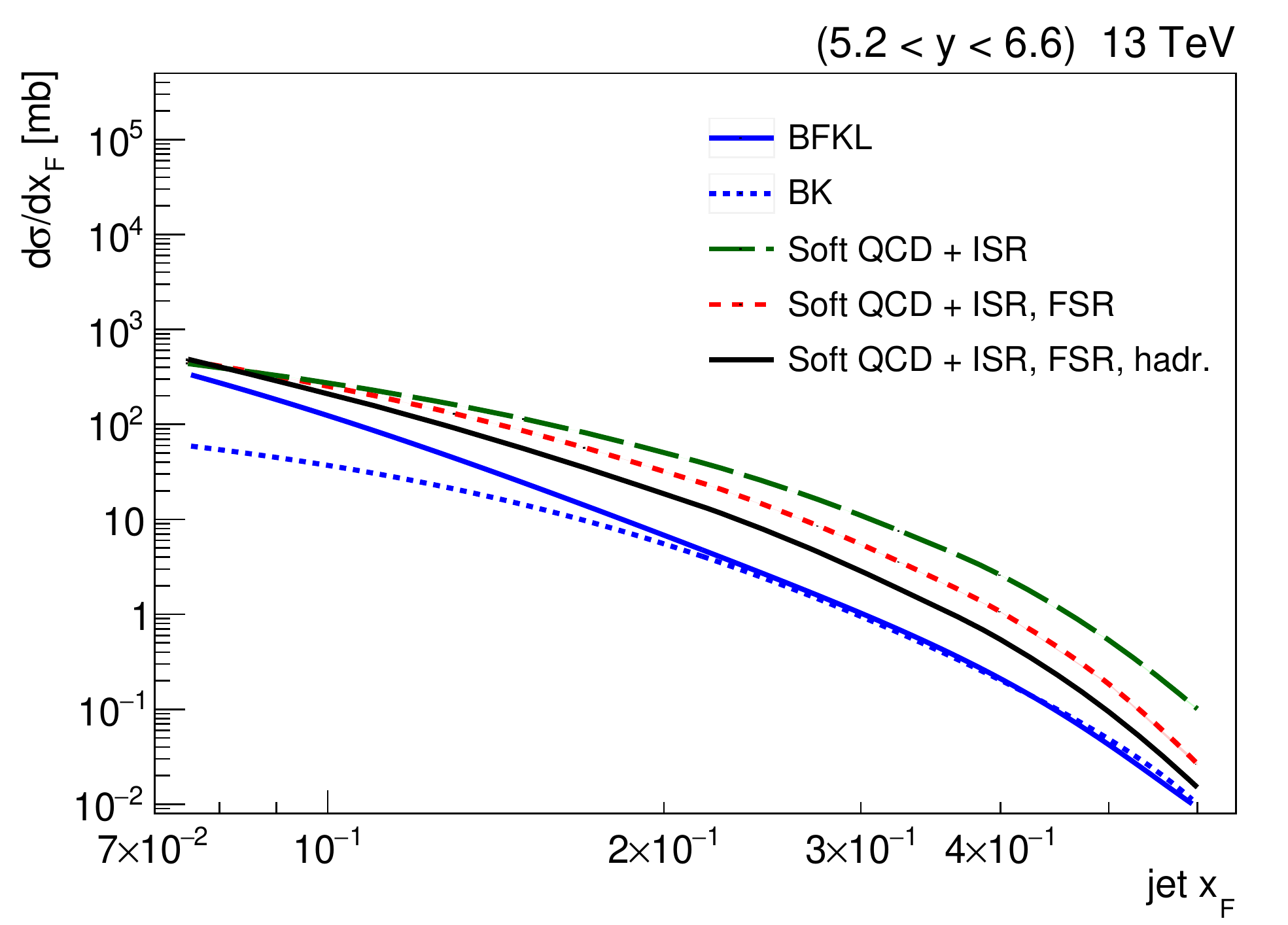}                
\caption{Differential jet cross sections as function of jet $\xf$ at forward rapidity $5.2 < y < 6.6$ and for $\sqrt{s} = 7$ TeV (left) and 13 TeV (right).  Calculations obtained with the HEF framework (labeled ``BFKL'' and ``BK'') are compared to simulations obtained with \pythia\ for soft QCD processes with ISR, adding subsequently FSR and hadronization.}
\label{fig:xFspectra}
\end{figure}

Similar conclusions can indeed be reached for the energy and $\xf$ spectra shown in Figs. \ref{fig:Energyspectra} and \ref{fig:xFspectra}, although some care in interpreting the results should be taken: in a fixed rapidity window, the highest-energy jets tend to be the most forward, while the highest-$\pt$ jets are most central.  This may explain different tendencies w.r.t.\@ the $\pt$ spectra.  

We note that, after all effects in \pythia\ are included, the resulting spectra become remarkably similar to the prediction obtained within the HEF framework with the linear evolution equation, especially for $\sqrt{s} = 7$ TeV.

\subsection{Ratio of cross sections at $\sqrt{s} =$ 13 and 7 TeV}

Measuring cross section ratios helps to substantially reduce experimental uncertainties, and may also help to disentangle various physical phenomena because some effects cancel in the ratio while others do not. 

\begin{figure}[t]
\centering
\includegraphics[width=0.49\textwidth]{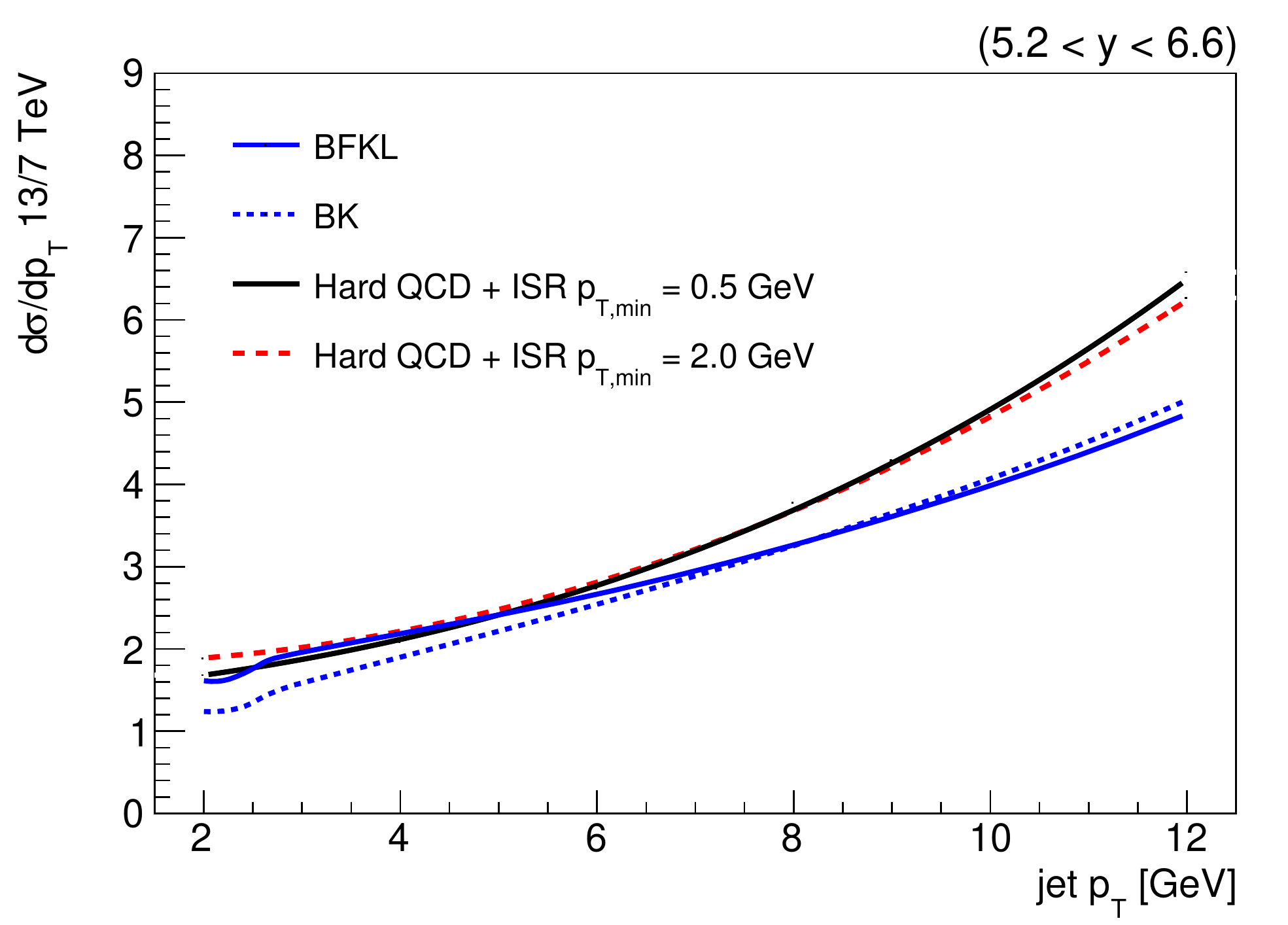}
\includegraphics[width=0.49\textwidth]{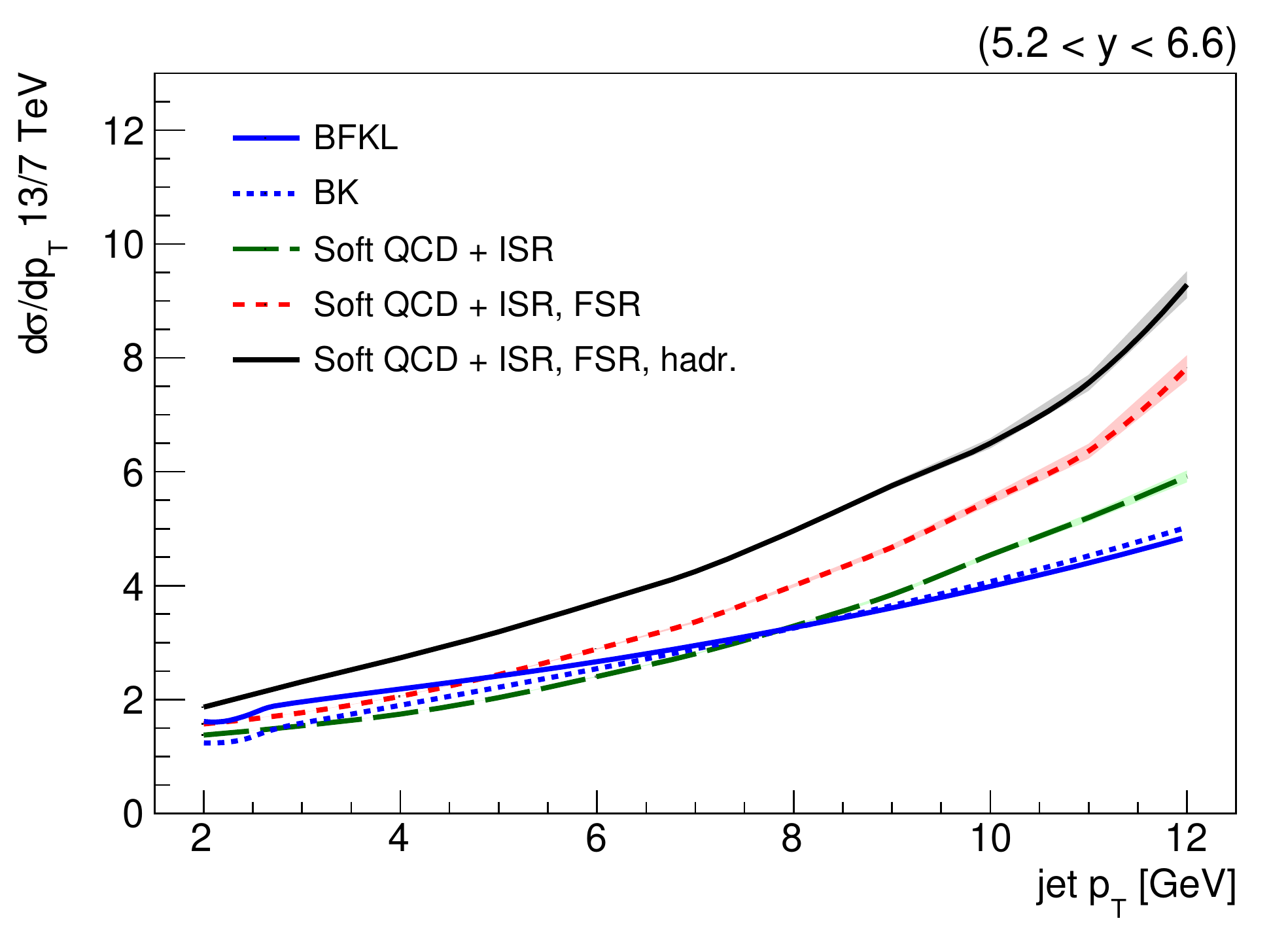}                
\caption{Ratio of the differential jet cross section at $\sqrt{s} = 13$ TeV w.r.t.\@ $\sqrt{s} = 7$ TeV as function of jet $\pt$ and at forward rapidity $5.2 < y < 6.6$.  Calculations obtained with the HEF framework (labeled ``BFKL'' and ``BK'') are compared to simulations obtained with \pythia\ for hard QCD processes with ISR (left) and soft QCD processes with ISR, FSR, and hadronization (right).}
\label{fig:pTratio}
\end{figure}

Figure~\ref{fig:pTratio} shows the ratio of the differential cross sections at $\sqrt{s} = 13$ and 7 TeV as function of jet $\pt$.  The cross section at $\sqrt{s} = 13$ TeV is larger than the cross section at $\sqrt{s} = 7$ TeV.  The ratio also increases with jet $\pt$, meaning that more hard jets are produced at the larger center-of-mass energy.  

As can be seen from both the HEF calculations and from the simulation of hard QCD events with \pythia\ (Fig.~\ref{fig:pTratio}, left), the effect of saturation will largely cancel in the cross section ratio.  Figure~\ref{fig:pTratio}, right, however shows that FSR and hadronization both result in a further increase of the cross section ratio.  This can be understood by the observation that FSR and hadronization result in a reduction of the cross section and that this reduction is stronger at $\sqrt{s} = 7$ TeV than for $\sqrt{s} = 13$ TeV (cf.\@ Fig.\@~\ref{fig:pTspectra_softQCD}).

\begin{figure}[t]
\centering
\includegraphics[width=0.49\textwidth]{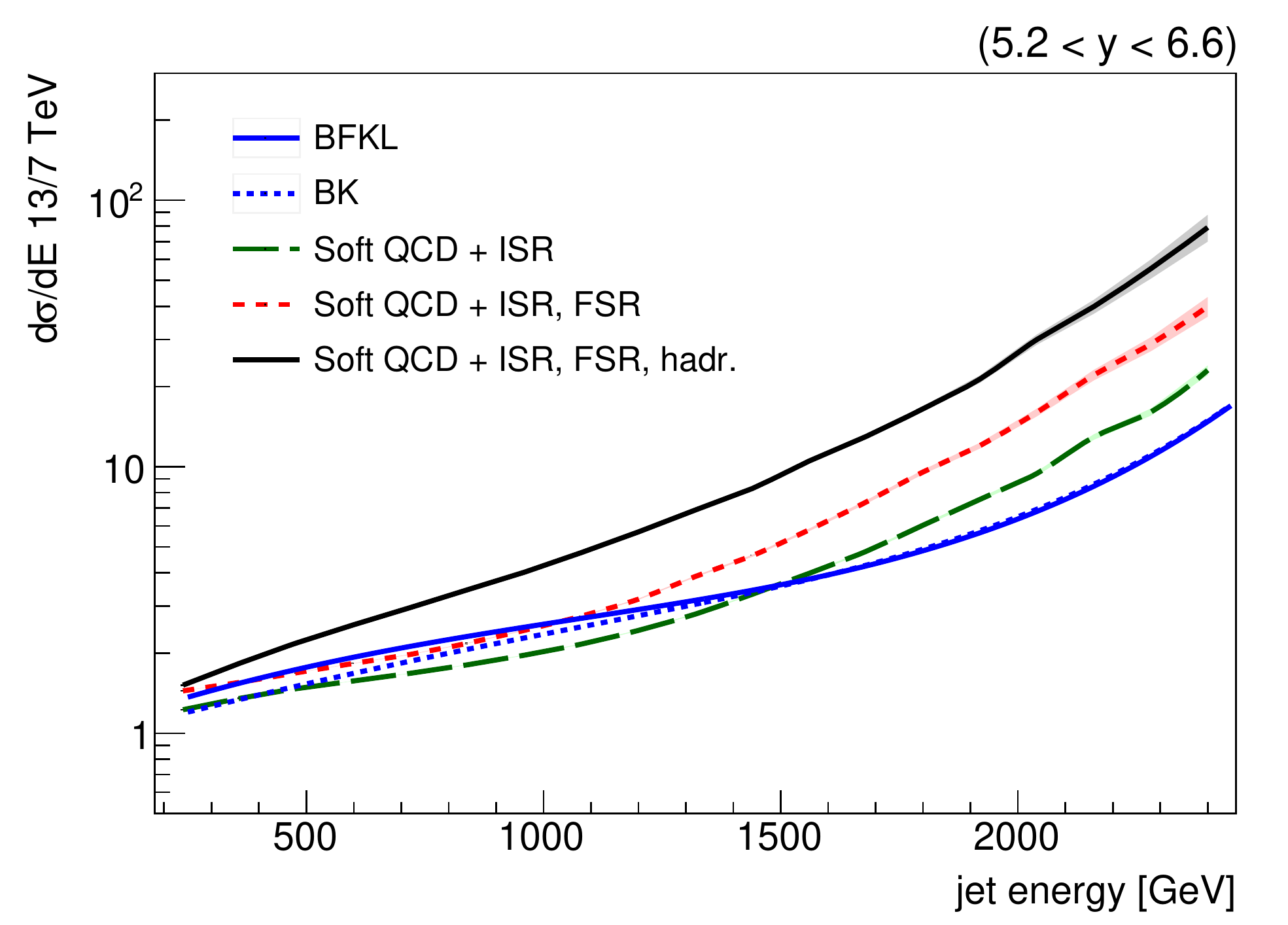}
\includegraphics[width=0.49\textwidth]{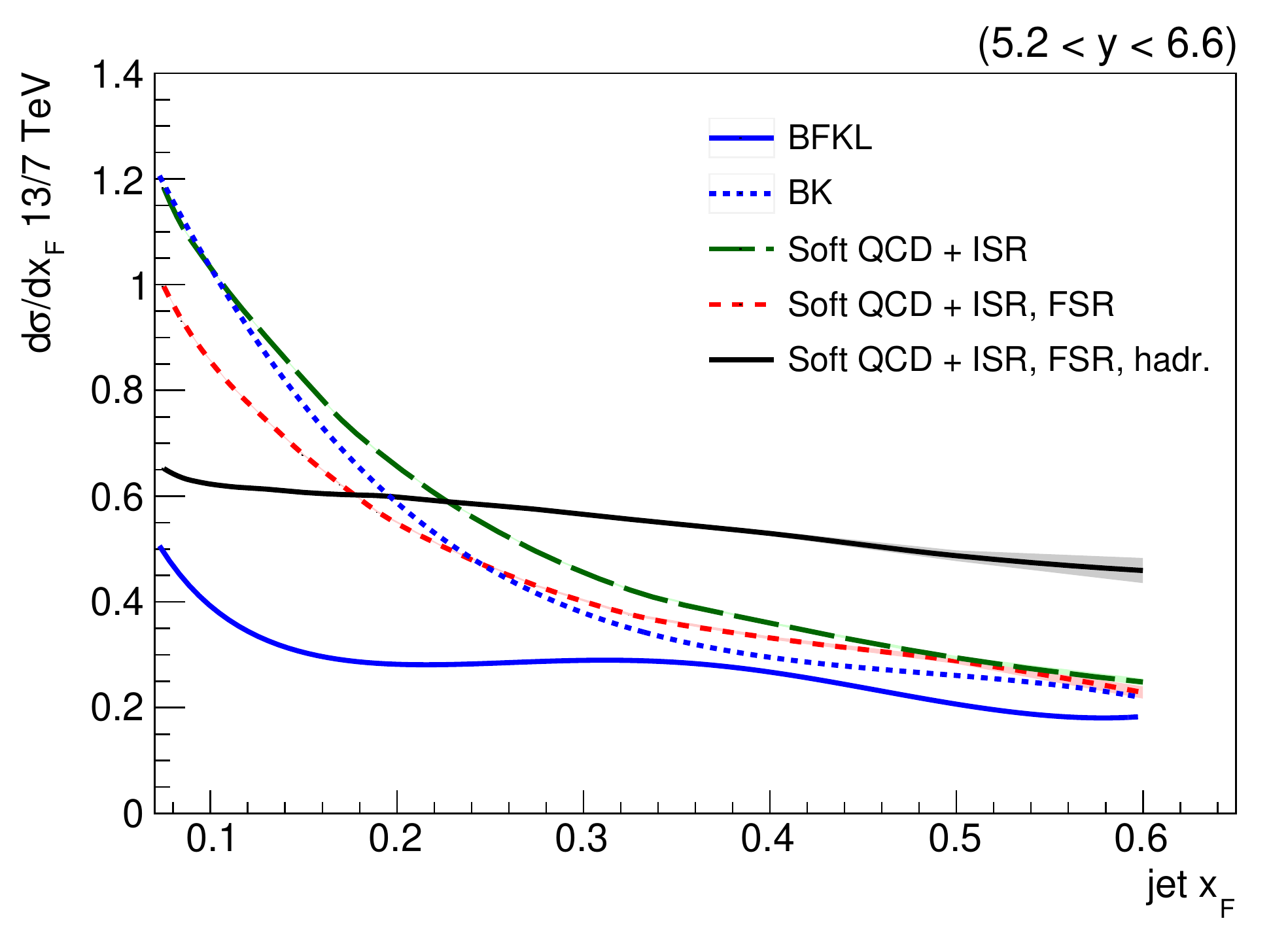}                
\caption{Ratio of the differential jet cross section at $\sqrt{s} = 13$ TeV w.r.t.\@ $\sqrt{s} = 7$ TeV as function of jet energy (left) and $\xf$ (right), at forward rapidity $5.2 < y < 6.6$.  Calculations obtained with the HEF framework (labeled ``BFKL'' and ``BK'') are compared to simulations obtained with \pythia \ for soft QCD processes with ISR, FSR, and hadronization.}
\label{fig:EnergyandxFratio}
\end{figure}

Similar conclusions can again be drawn from the cross section ratio as function of energy, as displayed in Fig.~\ref{fig:EnergyandxFratio} (left).  

The cross section ratio as function of $\xf$ is shown in Fig.~\ref{fig:EnergyandxFratio} (right).  The HEF calculation with linear evolution equation (BFKL) predicts a more or less constant cross section ratio around 0.2--0.3, with an increase at very low $\xf \sim 0.1$.  Both the HEF calculation with nonlinear evolution, and the \pythia\  simulations with regularization show a dramatic increase of the ratio below $\xf \sim 0.3$.  Of course, by taking the cross section ratio at a fixed value of $\xf$, one compares the jet cross section at very different values of $\pt$ or energy for the two center-of-mass energies.  For $\xf$ values below 0.3 this means that one compares the jet cross section at 7 TeV in the saturation/regularization domain to the jet cross section at 13 TeV at larger $\pt$, hence the sharp increase of the ratio.  Remarkably, this effect is smoothed out by hadronization, because the softening of the spectrum at larger $\pt$ by hadronization is less pronounced at $\sqrt{s} = 13$ TeV. The cross section ratio predicted by \pythia\ with all effects included is therefore again flat as a function of $\xf$, albeit at a larger value of around 0.5--0.6.

\section{Conclusions}
In this paper we provide predictions for various spectra characterizing single inclusive forward jet production in pp collisions at $\sqrt{s} = 7$ and 13 TeV.  Predictions obtained with the high-energy factorization (HEF) framework and $k_{\rm T}$-dependent parton densities are compared to simulations obtained with the \pythia\ Monte Carlo event generator.  Calculations in HEF based on the nonlinear BK evolution equations predict a suppression of the cross section at low $\pt$ and low energy scales relative to the ones based on the linear BFKL equation.  In \pythia\ this can be reproduced by introducing a lower cutoff on the transverse momentum exchange associated with the hard subprocess of $\hat{p}_{\rm T,min} = 2$ GeV. This is a strong indication that this $\pthat$ cutoff effectively introduces saturation effects in \pythia.  However, when introducing a more realistic regularization of the partonic cross section and multi-parton interactions, the \pythia\ prediction are closer to the HEF result with linear evolution. 
It has been shown that differences between the approaches discussed in this paper can be further elucidated by studying the ratio of cross sections at different center-of-mass energies, with $\pythia$ in general predicting a stronger increase and hardening of the cross section as compared to HEF. The ratio of the Feynman-$x$ observable between different center-of-mass energies in particular is very sensitive to saturation or regularization effects, and can substantially reduce experimental uncertainties.

\section*{Acknowledgements}

This research has been supported by a common FWO-PAS research grant.  The authors would like to thank Francesco Hautmann, Hannes Jung and Torbj\" orn Sj\" ostrand for useful discussions.


\begin{thebibliography}{99}


\bibitem{Gribov:1984tu}
  L.~V.~Gribov, E.~M.~Levin and M.~G.~Ryskin,
  Phys.\ Rept.\  {\bf 100} (1983) 1.


\bibitem{Balitsky:1995ub}
  I.~Balitsky,
  Nucl.\ Phys.\  {\bf B463 } (1996)  99-160.

\bibitem{Kovchegov:1999yj}
  Y.~V.~Kovchegov,
  Phys.\ Rev.\  D {\bf 60} (1999) 034008.

\bibitem{Kovchegov:1999ua}
  Y.~V.~Kovchegov,
  Phys.\ Rev.\  D {\bf 61} (2000) 074018.

\bibitem{JalilianMarian:1997gr}
  J.~Jalilian-Marian, A.~Kovner, A.~Leonidov and H.~Weigert,
  Phys.\ Rev.\ D {\bf 59} (1998) 014014
  doi:10.1103/PhysRevD.59.014014
  [hep-ph/9706377].

\bibitem{Iancu:2001ad}
  E.~Iancu, A.~Leonidov and L.~D.~McLerran,
  Phys.\ Lett.\ B {\bf 510} (2001) 133
  doi:10.1016/S0370-2693(01)00524-X
  [hep-ph/0102009].

\bibitem{GolecBiernat:1998js}
  K.~J.~Golec-Biernat and M.~Wusthoff,
  Phys.\ Rev.\ D {\bf 59} (1998) 014017
  doi:10.1103/PhysRevD.59.014017
  [hep-ph/9807513].

\bibitem{Albacete:2010pg}
  J.~L.~Albacete and C.~Marquet,
  Phys.\ Rev.\ Lett.\  {\bf 105} (2010) 162301
  doi:10.1103/PhysRevLett.105.162301
  [arXiv:1005.4065 [hep-ph]].

\bibitem{Dumitru:2010iy}
  A.~Dumitru, K.~Dusling, F.~Gelis, J.~Jalilian-Marian, T.~Lappi and R.~Venugopalan,
  Phys.\ Lett.\ B {\bf 697} (2011) 21
  doi:10.1016/j.physletb.2011.01.024
  [arXiv:1009.5295 [hep-ph]].


\bibitem{Dusling:2013qoz}
  K.~Dusling and R.~Venugopalan,
  Phys.\ Rev.\ D {\bf 87} (2013) no.9,  094034
  doi:10.1103/PhysRevD.87.094034
  [arXiv:1302.7018 [hep-ph]].



\bibitem{Kutak:2012rf}
  K.~Kutak and S.~Sapeta,
  Phys.\ Rev.\ D {\bf 86} (2012) 094043
  doi:10.1103/PhysRevD.86.094043
  [arXiv:1205.5035 [hep-ph]].

\bibitem{Chatrchyan:2008aa}
  CMS Collaboration,
  JINST {\bf 3} (2008) S08004.
  doi:10.1088/1748-0221/3/08/S08004

\bibitem{Catani:1990eg}
  S.~Catani, M.~Ciafaloni and F.~Hautmann,
  Nucl.\ Phys.\ B {\bf 366} (1991) 135.

\bibitem{Deak:2009xt}
  M.~Deak, F.~Hautmann, H.~Jung and K.~Kutak,
  JHEP {\bf 0909} (2009) 121
  doi:10.1088/1126-6708/2009/09/121
  [arXiv:0908.0538 [hep-ph]].


\bibitem{Sjostrand:2014zea}
  T.~Sjostrand {\it et al.},
  Comput.\ Phys.\ Commun.\  {\bf 191} (2015) 159
  doi:10.1016/j.cpc.2015.01.024
  [arXiv:1410.3012 [hep-ph]].

\bibitem{Sjostrand:2006za}
  T.~Sjostrand, S.~Mrenna and P.~Z.~Skands,
  JHEP {\bf 0605} (2006) 026
  doi:10.1088/1126-6708/2006/05/026
  [hep-ph/0603175].


\bibitem{Chirilli:2011km}
  G.~A.~Chirilli, B.~W.~Xiao and F.~Yuan,
  Phys.\ Rev.\ Lett.\  {\bf 108} (2012) 122301
  doi:10.1103/PhysRevLett.108.122301
  [arXiv:1112.1061 [hep-ph]].

\bibitem{Altinoluk:2014eka}
  T.~Altinoluk, N.~Armesto, G.~Beuf, A.~Kovner and M.~Lublinsky,
  Phys.\ Rev.\ D {\bf 91} (2015) no.9,  094016
  doi:10.1103/PhysRevD.91.094016
  [arXiv:1411.2869 [hep-ph]].

\bibitem{Iancu:2016vyg}
  E.~Iancu, A.~H.~Mueller and D.~N.~Triantafyllopoulos,
  JHEP {\bf 1612} (2016) 041
  doi:10.1007/JHEP12(2016)041
  [arXiv:1608.05293 [hep-ph]].

\bibitem{Stasto:2016wrf}
  A.~M.~Stasto and D.~Zaslavsky,
  Int.\ J.\ Mod.\ Phys.\ A {\bf 31} (2016) no.24,  1630039
  doi:10.1142/S0217751X16300398
  [arXiv:1608.02285 [hep-ph]].

\bibitem{Deak:2010gk}
  M.~Deak, F.~Hautmann, H.~Jung and K.~Kutak,
  arXiv:1012.6037 [hep-ph].

\bibitem{Deak:2011ga}
  M.~Deak, F.~Hautmann, H.~Jung and K.~Kutak,
  Eur.\ Phys.\ J.\ C {\bf 72} (2012) 1982
  doi:10.1140/epjc/s10052-012-1982-5
  [arXiv:1112.6354 [hep-ph]].

\bibitem{Deak:2009ae}
  M.~Deak, F.~Hautmann, H.~Jung and K.~Kutak,
  arXiv:0908.1870 [hep-ph].

\bibitem{Deak:2011gj}
  M.~Deak, F.~Hautmann, H.~Jung and K.~Kutak,
  arXiv:1112.6386 [hep-ph].



\bibitem{Kutak:2016mik}
  K.~Kutak, R.~Maciula, M.~Serino, A.~Szczurek and A.~van Hameren,
  JHEP {\bf 1604} (2016) 175
  doi:10.1007/JHEP04(2016)175
  [arXiv:1602.06814 [hep-ph]].

\bibitem{Kutak:2016ukc}
  K.~Kutak, R.~Maciula, M.~Serino, A.~Szczurek and A.~van Hameren,
  Phys.\ Rev.\ D {\bf 94} (2016) no.1,  014019
  doi:10.1103/PhysRevD.94.014019
  [arXiv:1605.08240 [hep-ph]].

\bibitem{Luszczak:2016csq}
  M.~Luszczak, R.~Maciula, A.~Szczurek and M.~Trzebinski,
  arXiv:1606.06528 [hep-ph].

\bibitem{Baranov:2015yea}
  S.~P.~Baranov, A.~V.~Lipatov and N.~P.~Zotov,
  Phys.\ Rev.\ D {\bf 93} (2016) no.9,  094012
  doi:10.1103/PhysRevD.93.094012
  [arXiv:1510.02411 [hep-ph]].

\bibitem{Sapeta:2015gee}
  S.~Sapeta,
  Prog.\ Part.\ Nucl.\ Phys.\  {\bf 89} (2016) 1
  doi:10.1016/j.ppnp.2016.02.002
  [arXiv:1511.09336 [hep-ph]].

\bibitem{Dooling:2014kia}
  S.~Dooling, F.~Hautmann and H.~Jung,
  Phys.\ Lett.\ B {\bf 736} (2014) 293
  doi:10.1016/j.physletb.2014.07.035
  [arXiv:1406.2994 [hep-ph]].



\bibitem{Dumitru:2005gt}
  A.~Dumitru, A.~Hayashigaki and J.~Jalilian-Marian,
  Nucl.\ Phys.\ A {\bf 765} (2006) 464
  doi:10.1016/j.nuclphysa.2005.11.014
  [hep-ph/0506308].

\bibitem{vanHameren:2012if}
  A.~van Hameren, P.~Kotko and K.~Kutak,
  JHEP {\bf 1301} (2013) 078
  doi:10.1007/JHEP01(2013)078
  [arXiv:1211.0961 [hep-ph]].




\bibitem{Bury:2016cue}
  M.~Bury, M.~Deak, K.~Kutak and S.~Sapeta,
  Phys.\ Lett.\ B {\bf 760} (2016) 594
  doi:10.1016/j.physletb.2016.07.041
  [arXiv:1604.01305 [hep-ph]].


\bibitem{Kuraev:1977fs}
  E.~A.~Kuraev, L.~N.~Lipatov and V.~S.~Fadin,
  Sov.\ Phys.\ JETP {\bf 45} (1977) 199
   [Zh.\ Eksp.\ Teor.\ Fiz.\  {\bf 72} (1977) 377].



\bibitem{Balitsky:1978ic}
  I.~I.~Balitsky and L.~N.~Lipatov,
  Sov.\ J.\ Nucl.\ Phys.\  {\bf 28} (1978) 822
   [Yad.\ Fiz.\  {\bf 28} (1978) 1597].


\bibitem{Kuraev:1976ge}
  E.~A.~Kuraev, L.~N.~Lipatov and V.~S.~Fadin,
  Sov.\ Phys.\ JETP {\bf 44} (1976) 443
   [Zh.\ Eksp.\ Teor.\ Fiz.\  {\bf 71} (1976) 840].


\bibitem{Skands:2014pea} 
  P.~Skands, S.~Carrazza and J.~Rojo,
  Eur.\ Phys.\ J.\ C {\bf 74}, no. 8, 3024 (2014)
  doi:10.1140/epjc/s10052-014-3024-y
  [arXiv:1404.5630 [hep-ph]].

\bibitem{Cacciari:2011ma} 
  M.~Cacciari, G.~P.~Salam and G.~Soyez,
  Eur.\ Phys.\ J.\ C {\bf 72}, 1896 (2012)
  doi:10.1140/epjc/s10052-012-1896-2
  [arXiv:1111.6097 [hep-ph]].

\bibitem{Cacciari:2008gp} 
  M.~Cacciari, G.~P.~Salam and G.~Soyez,
  JHEP {\bf 0804}, 063 (2008)
  doi:10.1088/1126-6708/2008/04/063
  [arXiv:0802.1189 [hep-ph]].

\bibitem{Jung}
Private communication with Hannes Jung.

\bibitem{Kotko:2016lej}
  P.~Kotko, A.~M.~Stasto and M.~Strikman,
  arXiv:1608.00523 [hep-ph].





\bibitem{Kwiecinski:1997ee}
  J.~Kwiecinski, A.~D.~Martin and A.~M.~Stasto,
  Phys.\ Rev.\ D {\bf 56} (1997) 3991
  doi:10.1103/PhysRevD.56.3991
  [hep-ph/9703445].





\end{thebibliography}
\end{document}